\documentclass[a4paper]{IEEEtran}
\usepackage{color}
\usepackage{leftidx}
\usepackage{cite}
\usepackage{microtype}
\usepackage{mathtools}
\usepackage{graphicx,amssymb,amstext,amsmath}
\usepackage{cleveref}
\usepackage[ruled,vlined,lined,ruled,linesnumbered]{algorithm2e}
\usepackage{float}
\usepackage{algorithmic}

\begin{document}

\title{Joint Beamforming and Power Control for Throughput Maximization in IRS-assisted MISO WPCNs}
\author{Yuan~Zheng,~Suzhi~Bi,~Ying-Jun~Angela~Zhang,~Xiaohui~Lin, and Hui~Wang\\
\thanks{Y.~Zheng, S.~Bi, X.~Lin, and H.~Wang are with the College of Electronics and Information Engineering, Shenzhen University, Shenzhen, China, 518060 (email: zhyu@szu.edu.cn; xhlin@szu.edu.cn; wanghsz@szu.edu.cn). S. Bi is also with the Peng Cheng Laboratory, Shenzhen, China, 518066.}
\thanks{Y. J. Zhang is with the Department of Information Engineering, The Chinese University of Hong Kong, Hong Kong (email: yjzhang@ie.cuhk.edu.hk).}}
\maketitle

\vspace{-10pt}
\maketitle
\begin{abstract}
   Intelligent reflecting surface (IRS) is an emerging technology to enhance the energy- and spectrum-efficiency of wireless powered communication networks (WPCNs). In this paper, we investigate an IRS-assisted multiuser multiple-input single-output (MISO) WPCN, where the single-antenna wireless devices (WDs) harvest wireless energy in the downlink (DL) and transmit their information simultaneously in the uplink (UL) to a common hybrid access point (HAP) equipped with multiple antennas. Our goal is to maximize the weighted sum rate (WSR) of all the energy-harvesting users. To make full use of the beamforming gain provided by both the HAP and the IRS, we jointly optimize the active beamforming of the HAP and the reflecting coefficients (passive beamforming) of the IRS in both DL and UL transmissions, as well as the transmit power of the WDs to mitigate the inter-user interference at the HAP. To tackle the challenging optimization problem, we first consider fixing the passive beamforming, and converting the remaining joint active beamforming and user transmit power control problem into an equivalent weighted minimum mean square error (WMMSE) problem, where we solve it using an efficient block-coordinate descent (BCD) method. Then, we fix the active beamforming and user transmit power, and optimize the passive beamforming coefficients of the IRS in both the DL and UL using a semidefinite relaxation (SDR) method. Accordingly, we apply a block-structured optimization (BSO) method to update the two sets of variables alternately. Numerical results show that the proposed joint optimization achieves significant performance gain over other representative benchmark methods and effectively improves the throughput performance in multiuser MISO WPCNs.

\end{abstract}

\vspace{-2ex}

\IEEEpeerreviewmaketitle

\section{Introduction}
With the advent of Internet-of-Things (IoT) era, tens of billions of wireless devices (WDs) are envisioned to be interconnected, which inevitably induce the explosion of mobile data traffic and the ever-growing demands for higher data rates. The demand for dramatic network capacity increase and ubiquitous connectivity in IoT networks boosts the research on promising wireless technologies, such as millimetre wave (mmWave), ultra-dense network (UDN) and massive multipleinput multiple-output (mMIMO) technologies \cite{2020:CL}. However, their advantageous communication performance often comes at a cost of high network energy consumption and/or hardware expense. To address this
problem, wireless powered communication networks (WPCNs) have been proposed \cite{2014:Bi,2018:Bi,2019:Bi} to use dedicated wireless energy transferring nodes to power the operation of communication devices. Compared with its conventional battery-powered counterpart, the WPCN has its advantages in lowering the operating cost and improving the robustness of communication service especially in low power applications, such as sensor and IoT networks. However, the major technical challenge in WPCNs lies in the low power transfer efficiency over long distance, resulting very limited harvested energy by the distributed WDs. Although several energy-efficient techniques, including user cooperation\cite{2014:Ju1}, ambient backscatter communication \cite{2019:Zheng}, multi-antenna technique \cite{2014:LL}, have been proposed to address this problem, the low energy transfer efficiency induced by the wireless channel attenuation is still a fundamental performance bottleneck of WPCN systems.

Recently, intelligent reflecting surface (IRS) technology has received widespread attentions of its application in wireless communications \cite{2020:HC}. In particular, an IRS comprises a massive number of reconfigurable reflecting elements and a smart controller. Each element reflects impinging electromagnetic waves with a controllable amplitude variation and phase shift using the IRS controller. By properly adjusting the reflecting elements of IRS, the reflected signals are coherently combined with those from the other paths at the receiver to maximize the signal strength. Compared to the use of conventional amplify-and-forward (AF) or decode-and-forward (DF) relay, IRS merely changes the end-to-end channel through passive reflection without amplifying or re-encoding the received signals. The recent advance in meta-surface technology \cite{2014:Tj} makes it feasible to reconfigure the reflecting coefficients in real time, thus greatly enhancing the applicability of IRS under wireless fading channel.
The integration of IRS technique in wireless communication network leads to many new technological innovations and networking paradigms.
In terms of the circuit implementations, practical IRS circuits include conventional reflect-arrays \cite{2019:WQQ1}, liquid crystal surfaces \cite{2017:S}, and software-defined meta-materials \cite{2018:CL}, among others. For new networking schemes, the utilization of IRS was extended to various communication scenarios, such as backscatter communication system\cite{2020:ZW}, cognitive radio network \cite{2019:YJ}, and the UAV-based communication scenario\cite{2020:LS}.

The essential advantage of deploying IRS lies in its ability to alter the wireless propagation environment to enhance the end-to-end channel strength in a passive and energy-efficient manner. This makes IRS a promising solution  to tackle the fundamental performance constraints of WPCN.  A large body of research on IRS-assisted WPCN has recently emerged in the literature \cite{2018:WQQ,2019:Huang,2019:Pan,2019:Cao,2020:Pan,2020:Guo,2020:Huang}. For instance, the authors in \cite{2018:WQQ} considered a joint design of active beamforming at the base station (BS) and passive beamforming at the IRS to minimize the total transmit power of the BS under the received user signal-to-noise ratio (SNR) constraints. \cite{2019:Huang} considered a downlink (DL) multiuser multiple-input single-output (MISO) scenario and maximized the energy efficiency of the BS by alternatively optimizing the transmit beamforming at the BS and the phase shifts at the IRS. Besides, the weighted sum-rate maximization problem for IRS-assisted system was investigated in various scenarios, e.g., MISO system \cite{2020:Guo}, multicell multiple-input multiple-output (MIMO) network \cite{2019:Pan}, and simultaneous wireless information and power transfer (SWIPT) system \cite{2020:Pan}. The authors in \cite{2019:Cao} proposed an IRS-assisted mmWAVE communication system in which the IRS is used to overcome the impact of blockage. \cite{2020:Huang} designed a deep reinforcement learning (DRL)-based algorithm to jointly optimize the active and passive beamforming in the IRS-aided system.

Most of the existing works adopt the IRS to assist either wireless energy transfer (WET) in the DL or wireless information transmission (WIT) in the UL. However, the UL and DL transmissions in WPCNs are highly correlated by the device energy causality. In this sense, a joint design of IRS-assisted DL and UL transmissions is needed to achieve the maximum communication performance in WPCNs. Although this joint design was recently studied in \cite{2020:Lyu}, it only considered the optimization of passive beamforming of the IRS in both the DL and UL. The major challenge resides in the joint design of the active beamforming of the HAP and the passive beamforming of the IRS in both DL and UL transmissions. Besides, the user transmit power is affected by both the DL energy transfer and the UL inter-user interference when spatial multiplexing is used.  However, to the best of our knowledge, this important research topic has not been studied so far.

\begin{figure}
  \centering
   \begin{center}
      \includegraphics[width=0.6\textwidth]{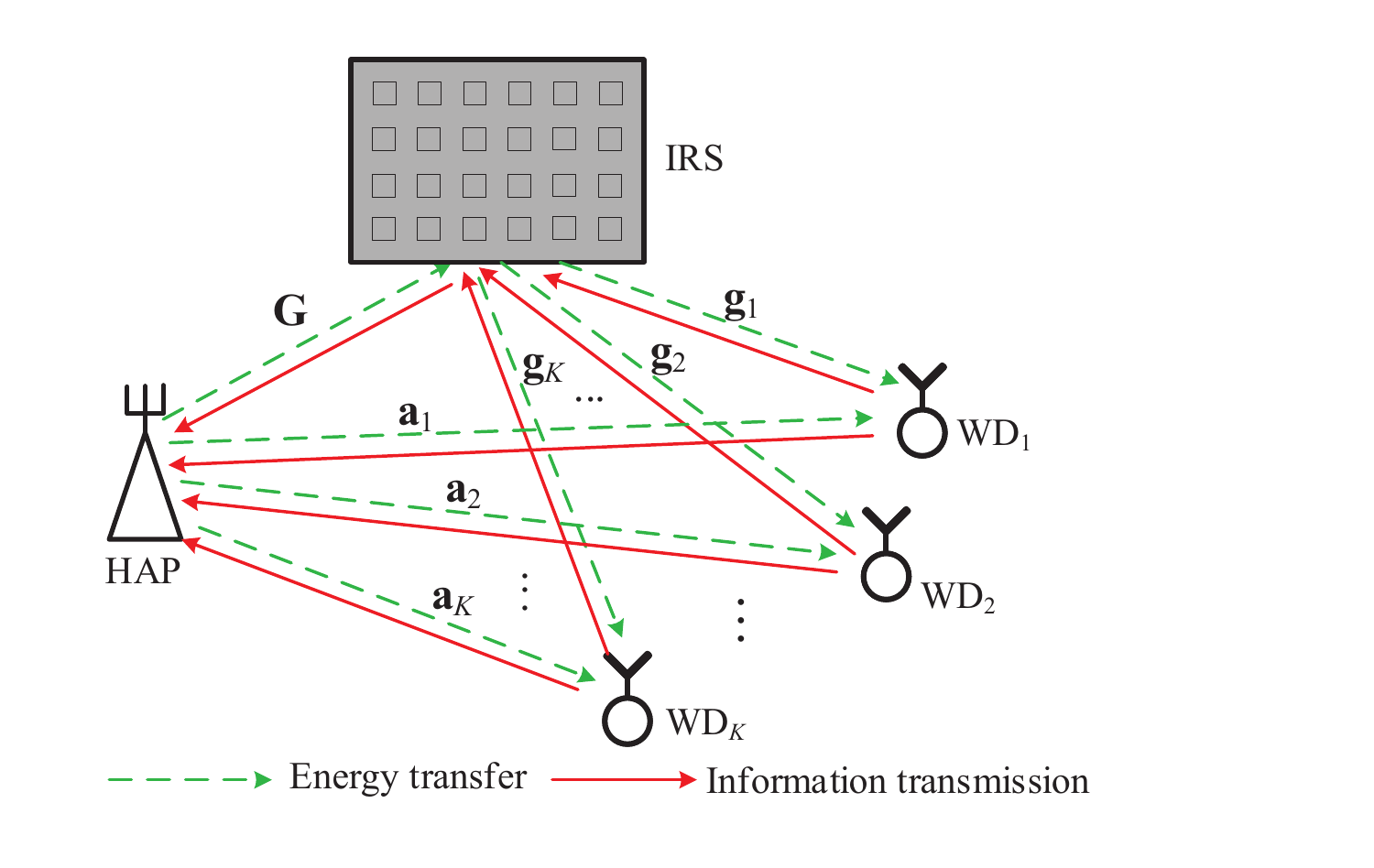}
   \end{center}\vspace{-.2in}
 \caption{The network structure of our proposed IRS-aided MISO WPCN.}
  \label{Fig.1}
\end{figure}

In this paper, we study the joint beamforming and user transmit power control problem in an IRS-assisted multiuser MISO WPCN. As shown in Fig.~\ref{Fig.1}, we consider a multi-antenna half-duplex HAP performing active beamforming to  broadcast wireless energy to all WDs in the DL and then receive information transmissions from the WDs in the UL. Specifically, the IRS performs passive beamforming by reflecting the transmitted energy (information) signals in the DL (UL) transmission. During the UL transmission, the WDs perform transmit power control to mitigate the multi-user interference at the HAP. Our objective is to maximize the weighted sum communication rate of all the WDs.
The main contributions of this paper are summarized as follows:
\begin{itemize}
  \item With the proposed IRS-assisted MISO WPCN, we first analyze the achievable data rates of all WDs. Then, we formulate an optimization problem to maximize the weighted sum rate (WSR) of all WDs by jointly optimizing the energy transmission time, the transmit power of the WDs, the active beamforming of the HAP and the passive beamforming the of IRS in both the UL and DL transmissions. The problem is highly non-convex because of the strong coupling of the design variables.
\item To tackle this non-convex problem, we first fix the energy transmission time and consider the joint beamforming and user transmit power control problem. Given passive beamforming of the IRS, we convert the remaining active beamforming and transmit power control problem into an equivalent weighted minimum mean square error (WMMSE) problem, which can be efficiently solved by applying a block-coordinate descent (BCD) method. In particular, we show that optimal energy beamforming matrix during the DL energy transfer of the HAP is rank-one and aligned to the maximum eigenmode of the weighted sum of DL channel matrices.

\item Given the active beamforming and user transmit power, we then propose a semidefinite relaxation (SDR) method to optimize the passive beamforming of the IRS, including the array reflecting coefficients in both DL and UL transmissions. Accordingly, we devise a block-structured optimization (BSO) technique to update the two sets of variables alternately. Finally, we apply a one-dimensional search method to obtain the optimal energy transmission time.
\end{itemize}

We conduct extensive simulations to evaluate the performance of the proposed IRS-assisted MISO WPCN. By comparing with the other representative benchmark methods, we show that the proposed method achieves a significant throughput performance gain in multiuser MISO WPCNs.

The rest of the paper is organized as follows: In Section II, we present the system model of the proposed IRS-assisted communication in multiuser MISO WPCN. We formulate the WSR optimization problem in Section III and propose an efficient algorithm to solve it in Section IV. In Section V, we perform simulations to evaluate the performance of the proposed method. Finally, Section VI concludes this paper.

\emph{Notations:} In this paper, vectors and matrices are denoted by boldface lowercase and uppercase letters, respectively. $\mathbb {C}^{m\times n}$ denotes the space of $m\times n$ complex-valued matrices. The operators $|\cdot|$, $\|\cdot\|$, $(\cdot)^T$ and $(\cdot)^H$ denote the absolute value, Euclidean norm, transpose and conjugate transpose, respectively. The symbols $\text{tr}(\mathbf X)$ and $\text{rank}(\mathbf X)$ denote the trace and rank of matrix $\mathbf X$, respectively. $\mathbb E[\cdot]$ stands for the statistical expectation. $\mathcal {CN}(\mu, \sigma^2)$ denotes the distribution of a circularly symmetric complex Gaussian (CSCG) random vector with mean $\mu$ and covariance $\sigma^2$. $\mathbf X\succeq 0$ means that $\mathbf X$ is positive semi-definite. $\arg(\cdot)$ denotes the phase extraction operation and $[\mathbf x]_{(1:N)}$ denotes the vector that contains the first $N$ elements of $\mathbf x$. $\text{diag}(\mathbf x)$ is a diagonal matrix withe the entries of the vector $\mathbf x$.

\section{System Model}
As show in Fig.~\ref{Fig.1}, we consider a multiuser MISO WPCN, which consists of one HAP and $K$ WDs. We define the set of WDs as $\mathcal {K}\triangleq\{1,\cdots,K\}$. It is assumed that the HAP is equipped with $M$ antennas and each WD has a single antenna. Specifically, the HAP broadcasts wireless energy to the WDs in the DL and receives wireless information transmission from the WDs in the UL.  All devices are assumed to operate over the same frequency band, where a time-division-duplexing (TDD) circuit is implemented at both the HAP and the WDs to separate the energy and information transmissions. The HAP performs energy beamforming in the DL and receive beamforming (e.g., MMSE) in the UL information transmission. The HAP has stable power supplies and each WD has an energy harvesting circuit and a rechargeable battery to store the harvested energy to power its operations. To enhance the propagation performance, we employ an IRS composed of
$N$ passive elements to assist the transmissions of the WPCN. The IRS can dynamically adjust the phase shift of each reflecting element based on the propagation environment \cite{2012:LP}. Due to the substantial path loss, we only consider one-time signal reflection by the IRS and ignore the signals that are reflected thereafter \cite{2018:WQQ}.

We assume that all channels follow a quasi-static flat fading model, where all the channel coefficients remain constant during each block transmission time, denoted by $T$, but vary from block to block. The baseband equivalent channels of HAP-to-IRS, IRS-to-WD$_i$, and HAP-to-WD$_i$ links are denoted as $\mathbf G\in \mathbb {C}^{M\times N}$, $\mathbf g_i\in \mathbb {C}^{N\times 1}$ and $\mathbf a_i\in \mathbb {C}^{M\times 1}, \forall i\in\mathcal K$, respectively. It is assumed that the channels of different transceiver pairs are independent to each other. Besides, the entries inside all channel vectors are modeled as zero-mean independent and identically distributed (i.i.d.) complex Gaussian random variables with variance depending on the path loss of the respective wireless links. The corresponding channel gains are denoted as $g_0=\|\mathbf G\|^2$, $g_i=\|\mathbf g_i\|^2$ and $h_i=\|\mathbf a_i\|^2$.
With the IRS-aided channel, each element at the IRS first combines all the received multi-path signals and then re-scatters the combined signal with a certain phase shift. Let $\boldsymbol\theta=[\theta_1,\theta_2,\cdots,\theta_N]$ and $\boldsymbol\Theta=\rm{diag}$$(\beta e^{j\theta_1},\cdots,\beta e^{j\theta_n},\cdots,\beta e^{j\theta_N})$
denote the phase-shift matrix of the IRS, where $\theta_n \in [0, 2\pi]$ and $\beta \in [0, 1]$ are the phase shift and amplitude reflection coefficient of each element, respectively. In this paper, we set $\beta=1$ for simplicity of in the following analysis, i.e., $\boldsymbol\Theta = \text{diag}(v_1, \cdots, v_n,\cdots, v_N)$ with $|v_n| = 1, n=1, \cdots, N$, and use transmit power control to mitigate the inter-user interference in the UL information transmission.

 As shown in the Fig.~2. we consider a harvest-then-transmit protocol that operates in two phases. In the first phase of duration $t$, the HAP transfers wireless energy in the DL for all the WDs to harvest. Meanwhile, the IRS scatters the incident signal from the HAP to the WDs, such that the WDs receive signals from both the direct-path and reflect-path channels. The remaining time of the block is assigned for the UL information transmission, during which WDs transmit their independent information to the HAP. Likewise, the IRS simultaneously scatters the signals transmitted by all WDs to the HAP.
\begin{figure}
  \centering
   \begin{center}
      \includegraphics[width=0.5\textwidth]{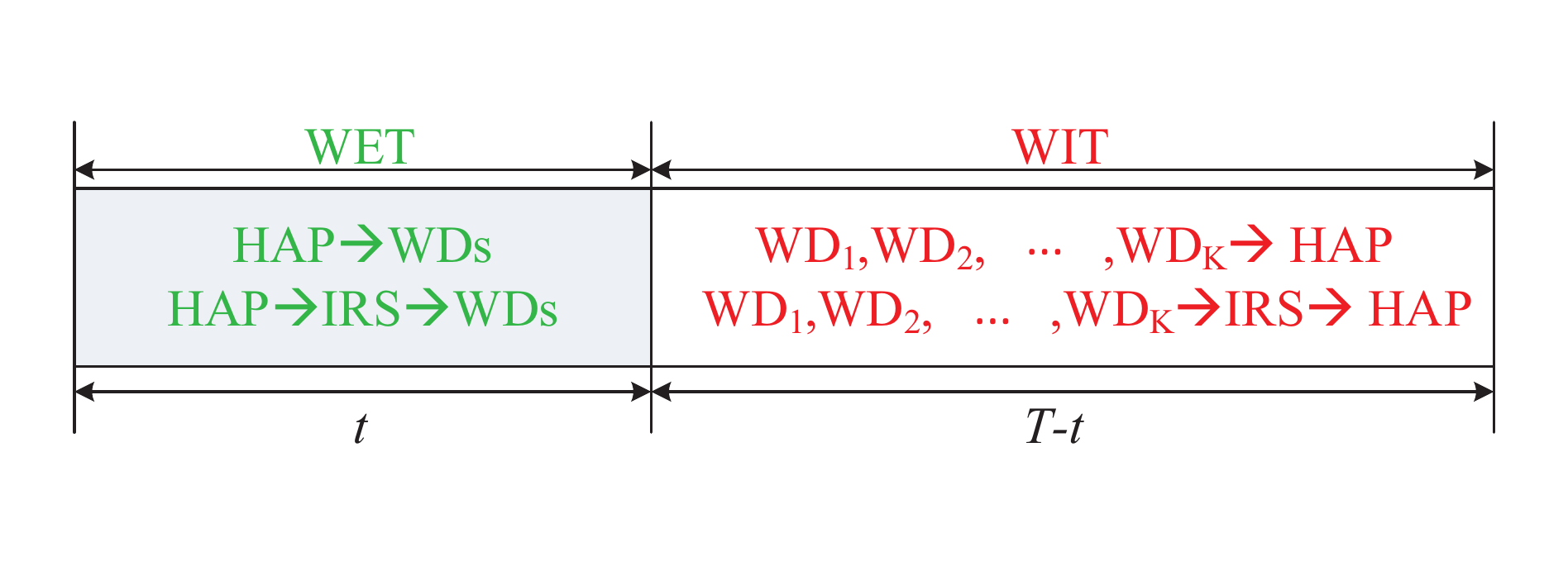}
   \end{center}\vspace{-.15in}
  \caption{The transmit protocol of the proposed IRS-assisted MISO WPCN.}
  \label{Fig.2}
\end{figure}

 We assume that the channel state information (CSI) of all channels are perfectly known at the HAP.\footnote{The CSI can be precisely estimated by the channel estimation methods for IRS system proposed in \cite{2019:DH} and \cite{2019:CJ}, which is out of the scope of this paper.} We jointly optimize the active beamforming of the HAP and passive beamforming of the IRS in both the UL and DL, the user transmit power, and the transmission time allocation between the UL and DL transmissions, to maximize the WSR of all the users. In the following, we formulate the WSR maximization problem and propose an efficient method to solve it.

\section{Problem Formulation}
In this section, we derive the throughput of each WD and formulate the WSR maximization problem.
\subsection{Phase I: Energy Transfer}
In the first WET stage of duration $t$, we denote $\mathbf x(t)\in \mathbb {C}^{M\times 1}$ as the pseudo-random baseband energy signal transmitted by the HAP \cite{2014:Bi}. The transmit power is constrained by
\begin{equation}
\label{pc}
\mathbb E\left[|\mathbf x(t)|^2\right]=\text{tr}\left(\mathbb E\left[\mathbf x(t) \mathbf x(t)^H\right]\right)\triangleq\text{tr}(\mathbf W)\le P_0.
\end{equation}
where $\mathbf W\succeq0$ is the energy beamforming matrix, and $P_0$ denotes the maximum transmit power.

Then, the received signal by the $i$-th WD is expressed as \cite{2018:WQQ}
 \begin{equation}
 y_i^{(1)}(t)=(\mathbf G\boldsymbol\Theta_1\mathbf g_i+\mathbf a_i)^T\mathbf x(t)+n_i(t), \forall i\in\mathcal K,
 \end{equation}
where $\boldsymbol\Theta_1=\text{diag} (v_{1,1},\cdots,v_{1,N})$ denotes the energy reflection coefficient matrix at the IRS with $v_{1,n}=e^{j\theta_{1,n}},n=1,\cdots,N$, which satisfies $|v_{1,n}|=1$. $n_i(t)$ denotes additive white Gaussian noise (AWGN) at the receiver with $n_i(t)\sim \mathcal {CN}(0,N_0)$.

By neglecting the noise power, the amount of energy harvested by the $i$-th WD is
\begin{equation}
\label{energy}
 E_i^{(1)}=\eta\text{tr}\Big((\mathbf G\boldsymbol\Theta_1\mathbf g_i+\mathbf a_i)(\mathbf G\boldsymbol\Theta_1\mathbf g_i+\mathbf a_i)^H\mathbf W\Big)t, \forall i\in\mathcal K,
 \end{equation}
 where $0<\eta<1$ denotes the fixed energy harvesting efficiency for all the WDs.\footnote{Although a single energy harvesting circuit exhibits non-linear energy harvesting property due to the saturation effect of circuit, it is shown in \cite{2019:Kang} and \cite{2019:Ma} that the non-linear effect is effectively rectified by using multiple energy harvesting circuits concatenated in parallel, resulting in a sufficiently large linear conversion region. }
 Accordingly, the residual energy of the $i$-th WD is
 \begin{equation}
 \label{econ}
 E_i=\min\ \{E_0+E_i^{(1)}, E_{max}\}, \forall i\in \mathcal K,
  \end{equation}
  where $E_0$ is the known residual energy at the beginning of the current time slot, and $E_{max}$ is the battery capacity.

 \subsection{Phase II: Information Transmission}
 In the subsequent WIT phase of duration $T-t$, all WDs transmit their independent information simultaneously to the HAP using the harvested energy in phase I. Meanwhile, the IRS reflects signal of the WDs to the HAP. Let $s_i(t)$ denote the information signal transmitted by the $i$-th WD with $E[|s_i(t)|^2]=1$, and $P_i$ denote the transmit power of WD$_i$, which is restricted by
 \begin{equation}
 \label{pcon}
 (T-t)P_i+E_i^{(2)}\le E_i, \forall i\in\mathcal K,
 \end{equation}
where $E_i^{(2)}\ge0$ denotes the fixed energy consumption of WD$_i$ within a transmission block, such as the data processing unit and passive circuitry power consumption.

 Then, the received signals at the HAP from the $i$-th WD in the UL is
 \begin{equation}
 \begin{aligned}
 \label{y0}
 \mathbf y_i^{(2)}(t)=(&\mathbf G\boldsymbol\Theta_2\mathbf g_i+\mathbf a_i)\sqrt{P_i}s_i(t)+\\&\sum_{j \in \mathcal {K}\setminus i}(\mathbf G\boldsymbol\Theta_2\mathbf g_j+\mathbf a_j)\sqrt{P_j}s_j(t)+\mathbf n_0(t),\forall i\in\mathcal K,
 \end{aligned}
 \end{equation}
where $\boldsymbol\Theta_2=\text{diag} (v_{2,1},\cdots,v_{2,N})$ denotes the reflection-coefficient matrix at the IRS with $|v_{2,n}|=1,n=1,\cdots,N$. $\mathbf n_0(t)\in \mathbb {C}^{M\times 1}$ denotes the AWGN vector at the HAP with $\mathbf n_0(t)\sim \mathcal {CN}(\mathbf 0,N_0\mathbf I)$.

It is assumed that the signals of different users are independent. In this paper, we consider linear receive beamforming at the HAP by treating the interference as noise. The estimated signal is expressed as
\begin{equation}
\hat s_i=\boldsymbol f_i^H \mathbf y_i^{(2)}(t),\forall i\in\mathcal K,
\end{equation}
where $\boldsymbol f_i\in \mathbb {C}^{M\times 1}$ denotes the receiver beamforming vector.

Then, the interference-plus-noise ratio (SINR) at the HAP for decoding the signal of the $i$-th WD is
\begin{equation}
\label{snr1}
\begin{aligned}
\gamma_i=\frac{\|\boldsymbol f_i^H(\mathbf G\boldsymbol\Theta_2\mathbf g_i+\mathbf a_i)\|^2P_i}{\sum_{j \in \mathcal {K}\setminus i}\|\boldsymbol f_i^H(\mathbf G\boldsymbol\Theta_2\mathbf g_j+\mathbf a_j)\|^2P_j+\|\boldsymbol f_i^H\|^2N_0},\\ \forall i\in\mathcal K.
\end{aligned}
\end{equation}
Thus, the achievable rate for information transmission of WD$_i$ in the UL is given by
\begin{equation}
\begin{aligned}
\label{ri}
R_i=\frac{T-t}{T}\log_2\left(1+\gamma_i\right), \forall i\in\mathcal K.
\end{aligned}
\end{equation}
where it is assumed without loss of generality that $T = 1$, such that $T$ is not present in the data rate expressions in the remainder of this paper.
Note that a trade-off exists in the WET duration $t$ to maximize $R_i$. Specifically, a larger $t$ leads to higher allowable transmit power but shorter remaining information transmission time.

\subsection{Problem Formulation}
In this paper, we focus on maximizing the WSR of the $K$ WDs by jointly optimizing the transmit time $t$, user transmit power $\mathbf P=[P_1,\cdots,P_K]$, the active beamforming of the HAP (including the energy beamforming matrix $\mathbf W$ and the receiver beamforming matrix $\mathbf F=[\boldsymbol f_1,\cdots,\boldsymbol f_K]$), and the passive beamforming of the IRS (i.e., the phase shift matrices $\boldsymbol\Theta_1$, $\boldsymbol\Theta_2$). Mathematically, it is formulated as
\begin{equation}
\begin{aligned}
\label{P1}
    (\rm{P1}): &~\max_{t, \mathbf P, \mathbf W,\mathbf F,\boldsymbol\Theta_1,\boldsymbol\Theta_2} & & \sum_{i\in\mathcal K} \omega_iR_i\\
    &~~~~~~~~\text{s. t.}  & & (\ref{pc}), (\ref{energy}), (\ref{econ})\ \text{and}\ (\ref{pcon}),\\
     & & &t,P_i\ge 0, \forall i\in\mathcal K,\\
    & & & |v_{i,n}|=1, i=1,2, n=1,\cdots,N.
\end{aligned}
\end{equation}
where $\omega_i\geq 0$ is the weighting factor controlling the scheduling priority of WD$_i$.

Notice that the objective function in (\ref{ri}) is not a concave function in the optimizing variables. Besides, due to the modulus constraints and the multiplicative terms of (\ref{energy}), (\ref{econ}) and (\ref{pcon}), (P1) is highly non-convex in its current form. In the next section, we first transform (P1) into an equivalent problem and propose an efficient optimization algorithm to solve it.

\section{Proposed Solution to (P1)}
We first fix $t = \bar{t}$ in (P1), partition the remaining optimization variables into two blocks, and alternatively optimize the two blocks of variables in an iterative manner \cite{2016:M}. Specifically, the optimization variables are partitioned as $\{\mathbf W,\mathbf F, \mathbf P\}$ and $\{\boldsymbol\Theta_1,\boldsymbol\Theta_2\}$. Then, we propose efficient algorithms to solve the joint active beamforming and transmit power control sub-problem (to optimize $\{\mathbf W,\mathbf F, \mathbf P\}$) and passive beamforming problem (to optimize $\{\boldsymbol\Theta_1,\boldsymbol\Theta_2\}$) separately in the following.

By assuming that the information messages of different users are independent, i.e., $E[s_is_j]=0$ if $i\neq j$, we write the mean-square error (MSE) as
\begin{equation}
\label{ei1}
\begin{aligned}
e_i&=\mathbb E_{s_i}\left\{|\hat s_i-s_i|^2\right\}\\&=|\boldsymbol f_i^H(\mathbf G\boldsymbol\Theta_2\mathbf g_i+\mathbf a_i)\sqrt{P_i}-1|^2\\&\ \ \ +\sum_{j \in \mathcal {K}\setminus i}|\boldsymbol f_i^H(\mathbf G\boldsymbol\Theta_2\mathbf g_j+\mathbf a_j)|^2P_j+N_0\|\boldsymbol f_i^H\|^2, \forall i\in\mathcal K.
\end{aligned}
\end{equation}
Following the celebrated rate-MMSE equivalence established in \cite{2011:Shi}, we show in the following proposition that the original maximization problem can be transformed to a more tractable optimization problem.

\textbf{Proposition 1}: Given $t = \bar{t}$, the WSR maximization problem is equivalent to the following WMMSE problem,
\begin{equation}
\begin{aligned}
\label{P21}
   (\rm{P2}): &\min_{\mathbf W, \mathbf F, \mathbf P,\boldsymbol q,\boldsymbol\Theta_1,\boldsymbol\Theta_2} & & \sum_{i\in\mathcal K}\omega_i(1-\bar t)\Big(q_ie_i-\log(q_i)-1\Big)~~~~\\
    &~~~~~~~\text{s. t.}  & &(\ref{pc}), (\ref{energy}), (\ref{econ})\ \text{and}\ (\ref{pcon}),\\
   & & & P_i,q_i\ge 0,\forall i\in\mathcal K,\\
   & & & |v_{i,n}|=1, i=1,2, n=1,\cdots,N,
\end{aligned}
\end{equation}
where $\boldsymbol q=[q_1,\cdots,q_K]$ and $q_i$ is a positive weight variable for $i=1\cdots,K$.

\emph{Proof}: Please refer to Appendix 1.

With the above transformation, we design an efficient alternating optimization algorithm to solve the active beamforming and transmit power control sub-problem described as follows.

\subsection{Optimizing the active beamforming matrices and transmit power $\{\mathbf W, \mathbf F,\mathbf P\}$}
 We first optimize $\{\mathbf W,\mathbf F,\mathbf P\}$ under fixed $\boldsymbol\Theta_1$ and $\boldsymbol\Theta_2$ . Define $\boldsymbol b_{i}=\mathbf G\boldsymbol\Theta_1\mathbf g_i+\mathbf a_i$ and $\tilde{\boldsymbol b}_{i}=\mathbf G\boldsymbol\Theta_2\mathbf g_i+\mathbf a_i, \forall i\in\mathcal K$. Then, (\ref{energy}) and (\ref{snr1}) are respectively expressed as,
  \begin{equation}
\label{energy2}
 E_i^{(1)}=\eta\text{tr}(\boldsymbol b_{i}\boldsymbol b_{i}^H\mathbf W)\bar t, \forall i\in\mathcal K,
 \end{equation}
 \begin{equation}
\label{snr2}
\begin{aligned}
\gamma_i=\frac{\|\boldsymbol f_i^H\tilde {\boldsymbol b}_i\|^2P_i}{\sum_{j \in \mathcal {K}\setminus i}\|\boldsymbol f_i^H\tilde {\boldsymbol b}_j\|^2P_j+\|\boldsymbol f_i^H\|^2N_0},\forall i\in\mathcal K.
\end{aligned}
\end{equation}

 Accordingly, we reformulate problem (P2) into the following equivalent problem
\begin{equation}
\begin{aligned}
\label{P21}
   ~~~~(\rm{P3}): &~~~\min_{\mathbf F,\mathbf W, \mathbf P,\boldsymbol q} & & \sum_{i\in\mathcal K}\omega_i(1-\bar t)\Big(q_ie_i-\log(q_i)-1\Big)~~~~~~~~~~~\\
    &~~~~~~\text{s. t.}  & &(\ref{pc}), (\ref{econ}), (\ref{pcon})\ \text{and}\ (\ref{energy2}),~~~~~~~~~~~~\\
   & & & P_i,q_i\ge 0,\forall i\in\mathcal K,
\end{aligned}
\end{equation}
 where $e_i$ is re-written as
 \begin{equation}
 \begin{aligned}
 \label{MSE}
e_i=|\boldsymbol f_i^H\tilde {\boldsymbol b}_i\sqrt{P_i}-1|^2+\sum_{j \in \mathcal {K}\setminus i}|\boldsymbol f_i^H\tilde {\boldsymbol b}_j|^2P_j+N_0\|\boldsymbol f_i^H\|^2.
\end{aligned}
 \end{equation}

 Note that the objective function of problem (P3) is convex  over each of the optimization variable $\boldsymbol f_i$, $P_i$ and $q_i$ for all $i\in\mathcal K$. Following \cite{2011:Shi}, we employ a block-coordinate descent (BCD) approach to tackle this problem. Specifically, we optimize one of the block variables in $\{\boldsymbol f_i, P_i,q_i\}$ with the other two fixed. To obtain some insights on the optimal solution structure, we apply the Lagrange duality method to solve (P3). The partial Lagrangian of problem (\ref{P21}) is formulated as
 \begin{equation}
\begin{aligned}
\mathcal{L}(\mathbf F,\mathbf W, \mathbf P,\boldsymbol q,\boldsymbol\mu)=&(1-\bar t)\sum_{i\in\mathcal K}\omega_i\Big(q_ie_i-\log(q_i)-1\Big)\\&+\sum_{i\in\mathcal K}\mu_i\left((1-\bar t)P_i+E_i^{(2)}\right)\\&-\mu_0 P_0-\text{tr}(\mathbf{AW}),
\end{aligned}
\end{equation}
 where $\mathbf A=\sum_{i\in\mathcal K}\mu_i\eta\bar t\boldsymbol b_i\boldsymbol b_i^H-\mu_0 \mathbf I$, $\mu_0$ and $\mu_i,\forall i\in\mathcal K$ are the nonnegative dual variables corresponding to the constraints (\ref{pc}) and (\ref{pcon}), respectively. For convenience, we denote $\boldsymbol \mu=[\mu_0,\mu_1,\cdots,\mu_K]$. Then, the dual function of (P3) is
\begin{equation}
      \label{dual}
     \begin{aligned}
    d(\boldsymbol\mu)= & ~\min_{\mathbf{F,W,P},\boldsymbol{q}} & &  \mathcal{L}(\mathbf{F,W,P},\boldsymbol{q,\mu}) \\
    &~~~~\text{s. t.}   & & \mathbf F,\mathbf W\succeq 0,\mathbf P,\boldsymbol{q}\geq 0.
    \end{aligned}
    \end{equation}
and the dual problem is
\begin{equation}
\begin{aligned}
\label{P3a}
   ~~~~(\rm{P3a}): &\max_{\boldsymbol\mu} && d(\boldsymbol\mu)\\
    &~\text{s. t.}  & &\boldsymbol\mu\ge0.
\end{aligned}
\end{equation}

Therefore, we first investigate the optimal solution of the dual function in (\ref{dual}) given a set of dual variables. Secondly, we determine the optimal dual variables $\mu_0^*$ and $\mu_i^*,\forall i\in\mathcal K$ to maximize the dual function.

\textbf{Proposition 2}: The optimal energy bemaforming matrix $\mathbf W^*$ for problem (P3) is expressed as
\begin{equation}
\label{W}
\mathbf W^*=P_0\mathbf u_1\mathbf u_1^H.
\end{equation}
where $\mathbf u_1$ is the unit-norm eigenvector of a matrix $\mathbf B=\sum_{i\in\mathcal K}\mu_i^*\eta\bar t\boldsymbol b_{i}\boldsymbol b_{i}^H$ corresponding to the maximum eigenvalue $\lambda_1$. Moreover, the optimal dual variables must satisfy $\mu_i^*>0$ for $i=1,\cdots,K$ and $\mu_0^*=\lambda_1$.

\emph{Proof}: Please refer to Appendix 2.

\emph{Remark 1}: Note that the optimal energy matrix in (\ref{W}) is rank-one such that transmitting a single energy stream is the optimal strategy for the DL energy transfer. Besides, $\mathbf u_1$ is aligned with the maximum eigenmode of matrix $\mathbf B$. 
 Accordingly, the optimal energy signal $\mathbf x(t)$ is determined as $\mathbf x(t)=\sqrt{P_0}\mathbf u_1x(t)$, where $x(t)$ denotes an arbitrary random scalar with unit variance.

 Furthermore, by checking the first-order optimality conditions for maximizing dual function with respect to $q_i$, $\boldsymbol f_i$ and $P_i$, respectively,
 we have
 \begin{equation}
\begin{aligned}
e_i-\frac{1}{q_i}=0,
\end{aligned}
\end{equation}
\begin{equation}
\begin{aligned}
\sum_{j\in \mathcal K}2\boldsymbol f_i^H\|\tilde{\boldsymbol b}_j\|^2P_j-2\tilde{\boldsymbol b}_i^H\sqrt{P_i}+2N_0\boldsymbol f_i^H=\mathbf 0,
\end{aligned}
\end{equation}
\begin{equation}
\begin{aligned}
\sum_{j\in \mathcal K}\omega_jq_j|\boldsymbol f_j^H\tilde{\boldsymbol b}_i|^2-\frac{\omega_iq_i\boldsymbol f_i^H\tilde{\boldsymbol b}_i}{\sqrt{P_i}}+\mu_i=0,
\end{aligned}
\end{equation}
  for all $i\in \mathcal K$. Then, we update each block variable in a closed-form manner given by
  \begin{equation}
\label{omega}
q_i^*=\frac{1}{e_i},\forall i\in\mathcal K,
\end{equation}
\begin{equation}
\label{f3}
\begin{aligned}
\boldsymbol f_i^*=\frac{\tilde {\boldsymbol b}_i\sqrt{P_i}}{\sum_{j\in\mathcal K}\|\tilde {\boldsymbol b}_j\|^2P_j+N_0}, \forall i\in\mathcal K,
\end{aligned}
\end{equation}
\begin{equation}
\label{PPP}
P_i^*=\left(\frac{\omega_iq_i\boldsymbol f_i^H\tilde {\boldsymbol b}_i}{\sum_{j\in\mathcal K}\omega_jq_j|\boldsymbol f_j^H\tilde {\boldsymbol b}_i|^2+\mu_i^*}\right)^{2}, \forall i\in\mathcal K.
\end{equation}

After solving the dual function, we obtain the optimal dual variables $\mu_i^*$ by sub-gradient based algorithms, e.g., the ellipsoid method. The subgradient of $d(\boldsymbol \lambda)$ is denoted as $\boldsymbol\varsigma=[\varsigma_1,\cdots,\varsigma_K]$, where
\begin{equation}
\label{sub}
\varsigma_i=(1-\bar t)P_i^*+E_i^{(2)}-\eta\text{tr}(\boldsymbol{b}_i\boldsymbol{b}_i^H\mathbf W^*)\bar t, \forall i\in\mathcal K.
\end{equation}

\begin{algorithm}
\footnotesize
 \KwIn{$P_0$, $\mathbf G$, $t$, $\boldsymbol \Theta_1$, $\boldsymbol \Theta_2$, $N_0$, \{$\mathbf g_i, h_i, \forall i\in\mathcal K\}$; }
 \KwOut{$\boldsymbol q^*, \mathbf P^*, \mathbf W^*, \mathbf F^*$;}
{ \textbf{Initialize}: $j \leftarrow 0$, $\boldsymbol\mu^{(0)}>0$, feasible $\boldsymbol q^{(0)}$, $\mathbf F^{(0)}$, $\mathbf P^{(0)}$;\\
  \Repeat{\rm{The optimal objective value of primal problem (P3) converges}}{
  Calculate $\mathbf W^{(j+1)}$ using (\ref{W}) with given $\boldsymbol\mu^{(j)}$;\\
  Calculate $e_i$ in (\ref{MSE}) with given $\boldsymbol f_i^{(j)}$ and $P_i^{(j)}$;\\
  Calculate $q_i^{(j+1)}$ using (\ref{omega}) with given $\boldsymbol f_i^{(j)}$ and $P_i^{(j)}$;\\
  Calculate $\boldsymbol f_i^{(j+1)}$ using (\ref{f3}) with given $ q_i^{(j+1)}$ and $P_i^{(j)}$;\\
  Calculate $P_i^{(j+1)}$ using (\ref{PPP}) with given $q_i^{(j+1)}$ and $\boldsymbol f_i^{(j+1)}$;\\
  Calculate the sub-gradient of $\boldsymbol \mu^{(j)}$ using (\ref{sub});\\
  Update $\boldsymbol \mu^{(j+1)}$  by using the ellipsoid method;\\
   $j \leftarrow j+1$;\\}}
\textbf{Return} $\{\boldsymbol q^*, \mathbf P^*, \mathbf W^*, \mathbf F^*\}$ as a solution to (P3).
\caption{Proposed BCD method for problem (P3)}
\label{alg1}
\end{algorithm}
The detailed description of the BCD method for problem (P3) is summarized in Algorithm 1.

\subsection{Optimizing the passive beamforming matrices \{$\boldsymbol \Theta_1$,$\boldsymbol \Theta_2$\}}
Now, we optimize the phase shift matrices $\{\boldsymbol \Theta_1,\boldsymbol \Theta_2\}$ given fixed $\{\mathbf W,\mathbf F, \mathbf P\}$. Let $\boldsymbol v_n=[v_{n,1},\cdots,v_{n,N}]^T,n =1,2$. Define $\boldsymbol \zeta_i=\mathbf G\text{diag}(\mathbf g_i^T)\in \mathbb {C}^{M\times N}, \forall i\in\mathcal K$. Then, we have
\begin{equation}
\begin{aligned}
\mathbf G\boldsymbol\Theta_n\mathbf g_i+\mathbf a_i&=\mathbf G\text{diag}(\mathbf g_i^T)\boldsymbol v_n+\mathbf a_i\\&=\boldsymbol \zeta_i\boldsymbol v_n+\mathbf a_i.
\end{aligned}
\end{equation}
To tackle the non-convex modulus constraint in (P2), we first define $\bar{\boldsymbol v}_n=\left[\begin{matrix}\boldsymbol v_n\\1\end{matrix}\right]\in \mathbb {C}^{(N+1)\times 1},n=1,2$, $\bar{\boldsymbol\zeta_i}=\left[\begin{matrix}\boldsymbol \zeta_i, \mathbf a_i\end{matrix}\right]\in \mathbb {C}^{M\times (N+1)}$ and $\mathbf V_n=\bar{\boldsymbol v}_n\bar{\boldsymbol v}_n^H\in \mathbb {C}^{(N+1)\times (N+1)}$. Thus, we have 
\begin{equation}
\begin{aligned}
 &\|\boldsymbol f_i^H(\mathbf G\boldsymbol\Theta_2\mathbf g_i+\mathbf a_i)\|^2=\|\boldsymbol f_i^H(\boldsymbol \zeta_i\boldsymbol v_2+\mathbf a_i)\|^2\\=&\|\boldsymbol f_i^H\bar{\boldsymbol\zeta_i}\bar{\boldsymbol v}_2\|^2=\text{tr}(\mathbf V_2\bar{\boldsymbol \zeta}_i^H\boldsymbol f_i\boldsymbol f_i^H\bar{\boldsymbol \zeta}_i).
 \end{aligned}
\end{equation}
 Accordingly, we rewrite (\ref{energy}) as
\begin{equation}
\label{e4}
E_i^{(1)}=\eta \bar t \text{tr}(\mathbf V_1\bar{\boldsymbol \zeta}_i^H\mathbf W\bar{\boldsymbol \zeta}_i), \forall i\in\mathcal K.
\end{equation}
Consider the following transformation
\begin{equation}
\begin{aligned}
 &|\boldsymbol f_i^H(\mathbf G\boldsymbol\Theta_2\mathbf g_i+\mathbf a_i)\sqrt{P_i}-1|^2\\=&|\boldsymbol f_i^H(\boldsymbol \zeta_i\boldsymbol v_2+\mathbf a_i)\sqrt{P_i}-1|^2\\=&|\boldsymbol f_i^H\boldsymbol \zeta_i\boldsymbol v_2\sqrt{P_i}+\boldsymbol f_i^H\mathbf a_i\sqrt{P_i}-1|^2\\=&\|\widehat{\boldsymbol\zeta_i}\bar{\boldsymbol v}_2\|^2=\text{tr}(\mathbf V_2{\boldsymbol \psi}_i),
\end{aligned}
\end{equation}
where $\widehat{\boldsymbol\zeta_i}=\left[\begin{matrix}\boldsymbol f_i^H\boldsymbol \zeta_i\sqrt{P_i}, \boldsymbol f_i^H\mathbf a_i\sqrt{P_i}-1\end{matrix}\right]\in \mathbb {C}^{1\times (N+1)}$ and ${\boldsymbol \psi}_i=\widehat{\boldsymbol\zeta_i}^H\widehat{\boldsymbol\zeta_i}\in \mathbb {C}^{(N+1)\times (N+1)}, \forall i\in\mathcal K$. Then, the MSE in (\ref{ei1}) is re-expressed as
\begin{equation}
\begin{aligned}
e_i&(\mathbf V)=|\boldsymbol f_i^H(\mathbf G\boldsymbol\Theta_2\mathbf g_i+\mathbf a_i)\sqrt{P_i}-1|^2\\&\ \ \ +\sum_{j \in \mathcal {K}\setminus i}|\boldsymbol f_i^H(\mathbf G\boldsymbol\Theta_2\mathbf g_j+\mathbf a_j)|^2P_j+N_0\|\boldsymbol f_i^H\|^2\\&=\|\widehat{\boldsymbol\zeta_i}\bar{\boldsymbol v}_2\|^2+\sum_{j \in \mathcal {K}\setminus i}\|\boldsymbol f_i^H\bar{\boldsymbol\zeta_j}\bar{\boldsymbol v}_2\|^2P_j+N_0\|\boldsymbol f_i^H\|^2\\&=\text{tr}(\mathbf V_2{\boldsymbol \psi}_i)+\sum_{j \in \mathcal {K}\setminus i}\text{tr}(\mathbf V_2\bar{\boldsymbol \zeta}_j^H\boldsymbol f_i\boldsymbol f_i^H\bar{\boldsymbol \zeta}_j)P_j+N_0\text{tr}(\boldsymbol f_i\boldsymbol f_i^H).
\end{aligned}
\end{equation}

Note that $[\mathbf V_i]_{n,n}=1,i=1,2,n=1,\cdots,N+1$ hold from the modulus constraint of $v_{i,n}$ ($[\mathbf X]_{m,n}$ denotes the element in the $m$-th row and $n$-th column of matrix $\mathbf X$). Besides, $\mathbf V_i$ must satisfy rank$(\mathbf V_i) = 1$. Thus, we rewrite problem (P2) as
\begin{equation}
\begin{aligned}
\label{P4}
 ~ (\rm{P4}): &~~\min_{\mathbf V_1, \mathbf V_2} & &  \sum_{i=1}^K\omega_i(1-\bar t)\Big(q_ie_i(\mathbf V)-\log(q_i)-1\Big)\\
    &~~~\text{s. t.}  & &(\ref{econ}), (\ref{pcon})\ \text{and}\ (\ref{e4}),\\
    & & & [\mathbf V_i]_{n,n}=1,i=1,2,n=1,\cdots,N+1,\\
    & & &  \text{rank}\ (\mathbf V_i) = 1, \mathbf V_i\succeq 0.
\end{aligned}
\end{equation}
Dropping the non-convex rank-one constraint and removing the terms irrelevant to $\mathbf V_1,\mathbf V_2$, we reduce prblem (P4) to
\begin{equation}
\begin{aligned}
\label{P5b}
   (\rm{P4a}): &~~\min_{\mathbf V_1,\mathbf V_2\succeq 0} & & \sum_{i=1}^K\omega_iq_i{e}_i(\mathbf V)\\
     &~~~~~\text{s. t.}  & & (\ref{econ}), \ (\ref{pcon})\  \text{and} \ (\ref{e4})~~~~~~~~~~~~\\
    & & & [\mathbf V_i]_{n,n}=1,i=1,2,n=1,\cdots,N+1.
\end{aligned}
\end{equation}

Note that problem (P4a) is a standard semidefinite programming (SDP) and it can be efficiently solved by the optimization tools such as CVX \cite{2004:Boyd}. Let's denote the optimal solution to problem (P4a) as $\{\mathbf V_1^*, \mathbf V_2^*\}$. Generally, the relaxed problem (P4a) may not yield a rank-one solution.  To recover $\boldsymbol v_i$ from $\mathbf V_i^*$ for $i=1,2$, we obtain the eigenvalue decomposition of $\mathbf V_i^*$ as $\mathbf V_i^*={\boldsymbol U_i}{\boldsymbol\Sigma_i} {\boldsymbol U_i}^H$, where ${\boldsymbol U_i}\in \mathbb {C}^{(N+1)\times (N+1)}$ and ${\boldsymbol \Sigma_i}\in \mathbb {C}^{(N+1)\times (N+1)}$ denote a unitary matrix and diagonal matrix, respectively. Then, we apply the standard Gaussian randomization method \cite{2010:LZQ} to obtain a suboptimal solution $\bar{\boldsymbol v}_i$, i.e., $\bar{\boldsymbol v}_i={\boldsymbol U_i}{\boldsymbol\Sigma}_i^{1/2}{\boldsymbol r_i}, i=1,2$, where ${\boldsymbol r_i}\in \mathbb {C}^{(N+1)\times 1}$ is a random vector generated from ${\boldsymbol r_i} \sim \mathcal {CN}(\boldsymbol 0 ,\boldsymbol I_{N+1})$. With many candidate solutions $\boldsymbol r_i$'s, we select the best one $\bar{\boldsymbol v}_i$ among all $\boldsymbol r_i$ which minimizes the objective of (P4a). Finally, we obtain $\boldsymbol v_i^*=e^{j\arg([\bar{\boldsymbol v}_i]_{(1:N)}/\bar {v}_{i,N+1})}$, the optimal $\boldsymbol \Theta_1^*$ and $\boldsymbol \Theta_2^*$ can be obtained from $\boldsymbol v_1^*$ and $\boldsymbol v_2^*$, respectively.
\begin{algorithm}
\footnotesize
 \KwIn{$P_0$,  $N$, $\mathbf G$, $t$, $N_0$, $\{\mathbf g_i, h_i, \forall i\in\mathcal K\}$;}
 \KwOut{$\mathbf P^*, \mathbf W^*, \mathbf F^*, \boldsymbol \Theta_1^*, \boldsymbol \Theta_2^*$;}
{ \textbf{Initialize}: $k \leftarrow 0$,  $t \leftarrow \bar t$,  $\boldsymbol \Theta_1^{(0)}$ and $\boldsymbol \Theta_2^{(0)}$;\\
  \Repeat{\rm{The optimal objective value of (P1) converges}}{
  Calculate $\mathbf W^{(k+1)}$, $\mathbf F^{(k+1)}$ and $\mathbf P^{(k+1)}$ from Algorithm 1;\\
 Update $\mathbf V_i^{(k+1)}$ by solving SDP in (\ref{P5b}) and recover $\boldsymbol v_i^{(k+1)}$ ($\boldsymbol{\Theta}_i^{(k+1)}$) from $\mathbf V_i^{(k+1)}$; \\
  $k \leftarrow k+1$;}}
\textbf{Return} $\{\mathbf P^*, \mathbf W^*, \mathbf F^*, \boldsymbol \Theta_1^*, \boldsymbol \Theta_2^*\}$ as a solution to (P1).
\caption{Proposed BSO-based alternating iterative algorithm to problem (P1) with given $t=\bar t$}
\label{alg1}
\end{algorithm}

Based on the solutions to the two sub-problems (P3) and (P4), we devise an efficient iterative algorithm summarized in Algorithm 2. Specifically, given $t=\bar t$, the algorithm starts with certain feasible values of $\boldsymbol{\Theta}_1^{(0)}$ and $\boldsymbol{\Theta}_2^{(0)}$. Next, given a fixed solution $\{ \boldsymbol{\Theta}_1^{(k)}, \boldsymbol{\Theta}_2^{(k)}\}$ in the $k$-th iteration, we first obtain the optimal $\mathbf W^{(k+1)}$, $\mathbf F^{(k+1)}$ and $\mathbf P^{(k+1)}$ from Algorithm 1. Then, we update the phase shift matrices $\boldsymbol{\Theta}_1^{(k+1)}$ and $\boldsymbol{\Theta}_2^{(k+1)}$ using the SDR technique to solve problem (P4a) in the $(k+1)$-th iteration. The process repeats until convergence. At last, we obtain the optimal energy transmission time $t^*$ via a simple one-dimensional search method over $t\in(0,1)$, e.g., golden-section search \cite{2013:Ts} or the data-driven-based search \cite{2017:HZ}, which is omitted here for brevity.

\subsection{Convergence and Complexity Analysis}
The proposed BSO algorithm alternatingly solves two sub-problems (P3) and (P4) that optimize $\{\mathbf{W,F,P}\}$ and $\{\boldsymbol\Theta_1,\boldsymbol\Theta_2\}$, respectively. Following the Theorem 3 in \cite{2011:Shi}, the BCD method used in Algorithm 1 converges and the objective value of (P3) after optimization is non-increasing compared to that achieved by the initial input parameter. Besides, by our design, the randomization method used to solve (P4a) also guarantees that the objective is non-increasing after optimization. Due to the equivalence in \textbf{Proposition 1}, now that the objective of (P1) is non-decreasing in both the alternating steps and the optimal value of (P1) is bounded above, we conclude that the proposed Algorithm 2 converges asymptotically. In a practical setup, we will show the number of alternating iterations consumed by Algorithm 2 until convergence in simulation section.

We then analyze the complexity of Algorithm 2.  Here, we consider $N>M\geq K$ in a practical IRS-assisted multiuser MISO WPCN. The complexity of problem (P3) is dominated by the calculation of $\mathbf W^*$, which requires calculating the eigenvalue decomposition of an $M\times M$ matrix $\mathbf B$ with complexity of  $\mathcal O(M^3)$ \cite{2009:IM}. The SDP problem (P4a) can be solved with a worst-case complexity of $\mathcal O((N+1)^{4.5})$ \cite{2010:ZQ}. As we will show later in Fig.~9, the number of alternating iterations used by Algorithm 2 until convergence is of constant order, i.e., $\mathcal O(1)$, regardless of the value of $M$ and $N$. Therefore, the overall complexity of Algorithm 2 is $\mathcal O(M^3+(N+1)^{4.5})$.

\section{Simulation Results}
\begin{figure}[h]
  \centering
   \begin{center}
      \includegraphics[width=0.55\textwidth]{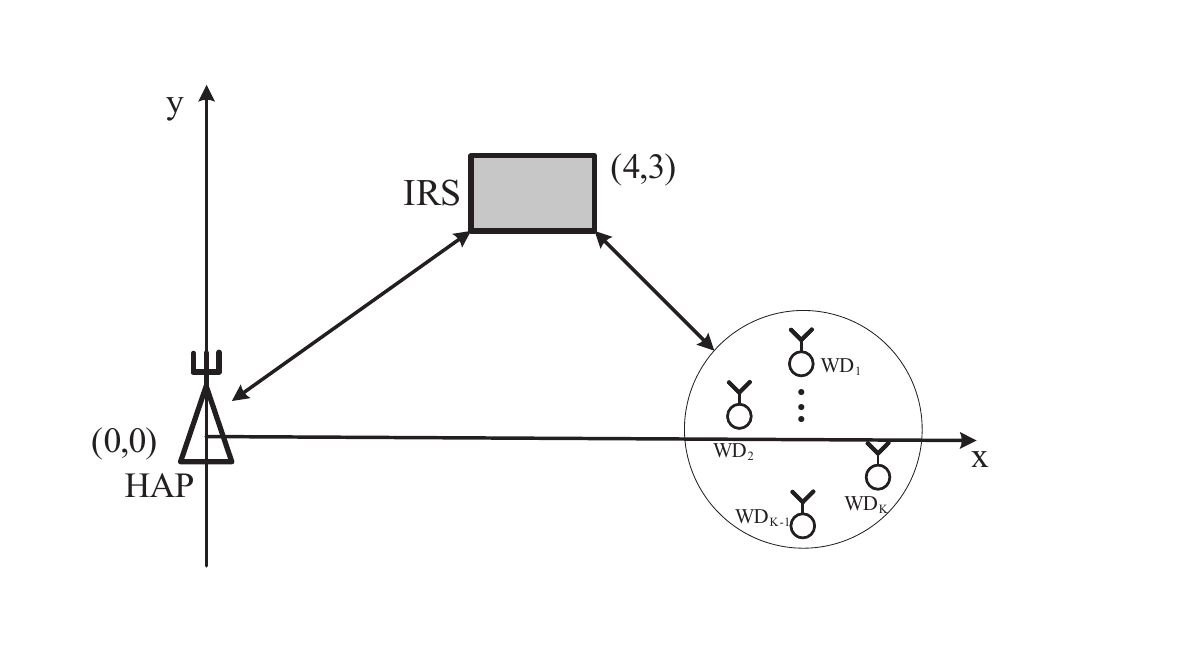}
   \end{center}
  \caption{The placement model of simulation setup.}
  \label{Fig.3}
\end{figure}

In this section, we provide numerical results to evaluate the performance of the proposed IRS-assisted MISO WPCN. In all simulations, we consider a two-dimensional (2D) coordinate system as shown in Fig.~\ref{Fig.3}, where the HAP and IRS are located at $(0,0)$ and $(4,3)$, the WDs are uniformly and randomly placed in a circle centered at $(d_c,0)$ with radius equal to 2 m \cite{2020:Lyu}. 
To account for the small-scale fading, we assume that all channels follow Rayleigh fading and the distance-dependent path loss is modeled as $L=C_0(\frac{d}{d_0})^{-\alpha}$, where $C_0$ is the constant path loss at the reference distance $d_0$, $d$ denotes the link distance, and $\alpha$ denotes the path loss exponent. To account for the heterogeneous channel conditions and avoid severe signal blockage, we set different path loss exponents of the HAP-IRS, IRS-WD$_i$ and HAP-WD$_i$ channels as $2.0, 2.2$ and $3.5$, respectively. For simplicity, we assume equal weights $\omega_i=1$ in all simulations.

\begin{table}[h]
\caption{System Parameters}
\label{tab:system parameters}
\centering
\begin{tabular}{|c|c|c|}
\hline
\footnotesize{\textbf{Parameter}} & \footnotesize{\textbf{Description}} & \footnotesize{\textbf{Value}}\\
\hline
$P_0$ & \footnotesize{Maximum transmission power of HAP} & $30$ dBm \\
\hline
$\eta$ & \footnotesize{Energy harvesting efficiency} & $0.8$ \\
\hline
$C_0$ & \footnotesize{Fixed path loss at reference distance} & $20$ dB \\
\hline
$N_0$ & \footnotesize{Noise power at receiver antenna} & $-90$ dBm \\
\hline
$K$&\footnotesize{Number of WDs}&4\\
\hline
$M$ & \footnotesize{Number of HAP antennas} & 6 \\
\hline
$N$ & \footnotesize{Number of reflecting elements} & 30 \\
\hline
$d_c$ & \footnotesize{Distance between the HAP and WDs} & $8$ m \\
\hline
$E_i^{(2)}$ & \footnotesize{Circuit energy consumption  of WD$_i$} & $10^{-6}$ J\cite{2014:LB}  \\
\hline
$\omega_i$ & \footnotesize{Weight factor of WD$_i$} & $1$  \\
\hline
\end{tabular}
\end{table}

Unless otherwise stated, the parameters used in the simulations are listed in Table I, which corresponds to a typical outdoor wireless powered sensor network similar to the setups in \cite{2020:Guo} and \cite{2020:Lyu}. The number of random vector for Gaussian randomization is set as 100 and the stopping criteria for the proposed algorithm is set as $10^{-4}$. All the simulation results are obtained by averaging over 1000 independent channel realizations.

In addition, we select three representative benchmark methods for performance comparison:
 \begin{enumerate}
  \item \emph{Passive beamforming optimization (PBO)}: We set uniform energy beamforming (i.e., $\mathbf W =\mathbf I_M$) and MMSE receive beamforming in line 3 of Algorithm 2. Then, the user transmit power $P_i$ and passive beamforming of the IRS $\{\boldsymbol\Theta_1,\boldsymbol\Theta_2\}$ are optimized alternatively in an iterative manner similarly to our proposed method. This method corresponds to the case that optimizes only the passive beamforming of the considered WPCN.
  \item \emph{Active beamforming optimization (ABO)}: In this case, the phase shifts of all reflecting elements at the IRS for both WET and
   WIT are fixed and uniformly generated as $\theta_{i,n}\in [0, 2\pi]$. The other variables are optimized using our proposed method. This method corresponds to the case where only the active beamforming of the HAP is optimized.
  \item \emph{Without IRS}: All WDs first harvest energy from the HAP and then transmit independently to the HAP. This corresponds to the method in \cite{2014:LL}.
 \end{enumerate}
 For fair comparison, we optimize the resource allocations in all the benchmark schemes. The details are omitted due to the page limit.

 \begin{figure}
  \centering
   \begin{center}
      \includegraphics[width=0.48\textwidth]{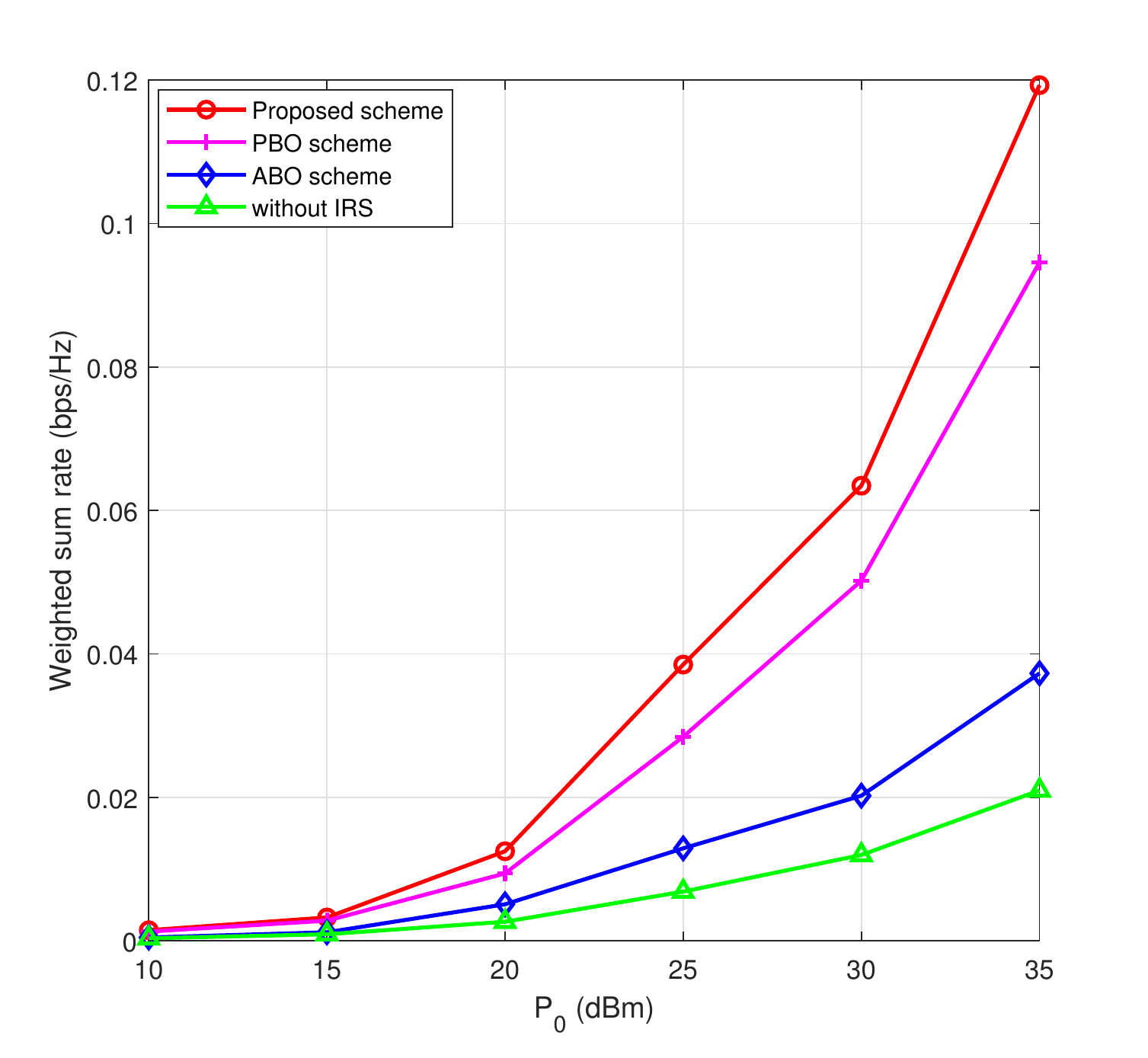}
   \end{center}\vspace{-.2in}
  \caption{The WSR performance versus the maximum transmit power of the HAP.}
  \label{Fig.4}
\end{figure}
\begin{figure}
  \centering
   \begin{center}
      \includegraphics[width=0.48\textwidth]{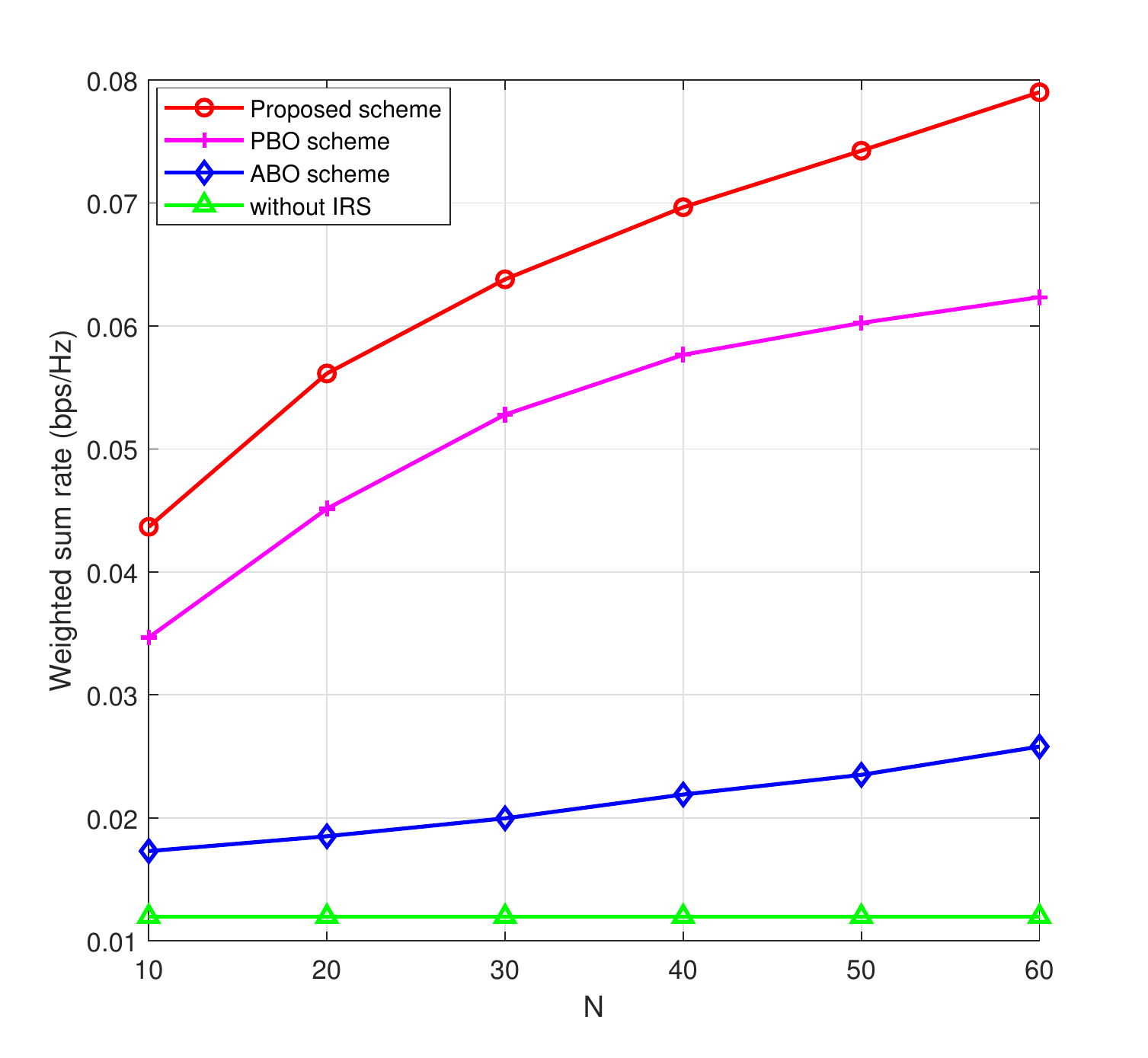}
   \end{center}\vspace{-.2in}
  \caption{The WSR performance versus the number of IRS reflecting elements.}
  \label{Fig.5}
\end{figure}
Fig.~\ref{Fig.4} shows the impact of the maximum transmit power of the HAP (i.e., $P_0$) to the WSR performance. As excepted, the WSR of all schemes increases with $P_0$ because the WDs are able to harvest more energy when the transmit power of HAP is higher. The joint optimization method achieves evident performance advantage over the other methods. In particular, the performance gap between the proposed scheme with the benchmark methods increases with $P_0$, which demonstrates its efficient usage of the harvested energy. It is also worth mentioning that even the IRS-assisted method with fixed phase shifts achieves better performance than that without the IRS thanks to the array energy gain provided by the IRS.

 In Fig.~\ref{Fig.5}, we study the impact of number of reflecting elements $N$ on the WSR performance when the value of $N$ varies from 10 to 60. We observe an evident increase of the WSR for the three IRS-assisted methods. In particular, compared to the ABO scheme, the slope of increase is larger for the proposed scheme and the PBO methods, because they can achieve extra beamforming gain besides the array gain of the IRS. By jointly optimizing the active and passive beamforming, our proposed scheme significantly outperforms the PBO and ABO schemes.  On average, the proposed joint optimization method achieves $23.66\%$, $205.57\%$ and $360.98\%$ higher throughput than the three benchmark methods, respectively.
\begin{figure}
  \centering
   \begin{center}
      \includegraphics[width=0.48\textwidth]{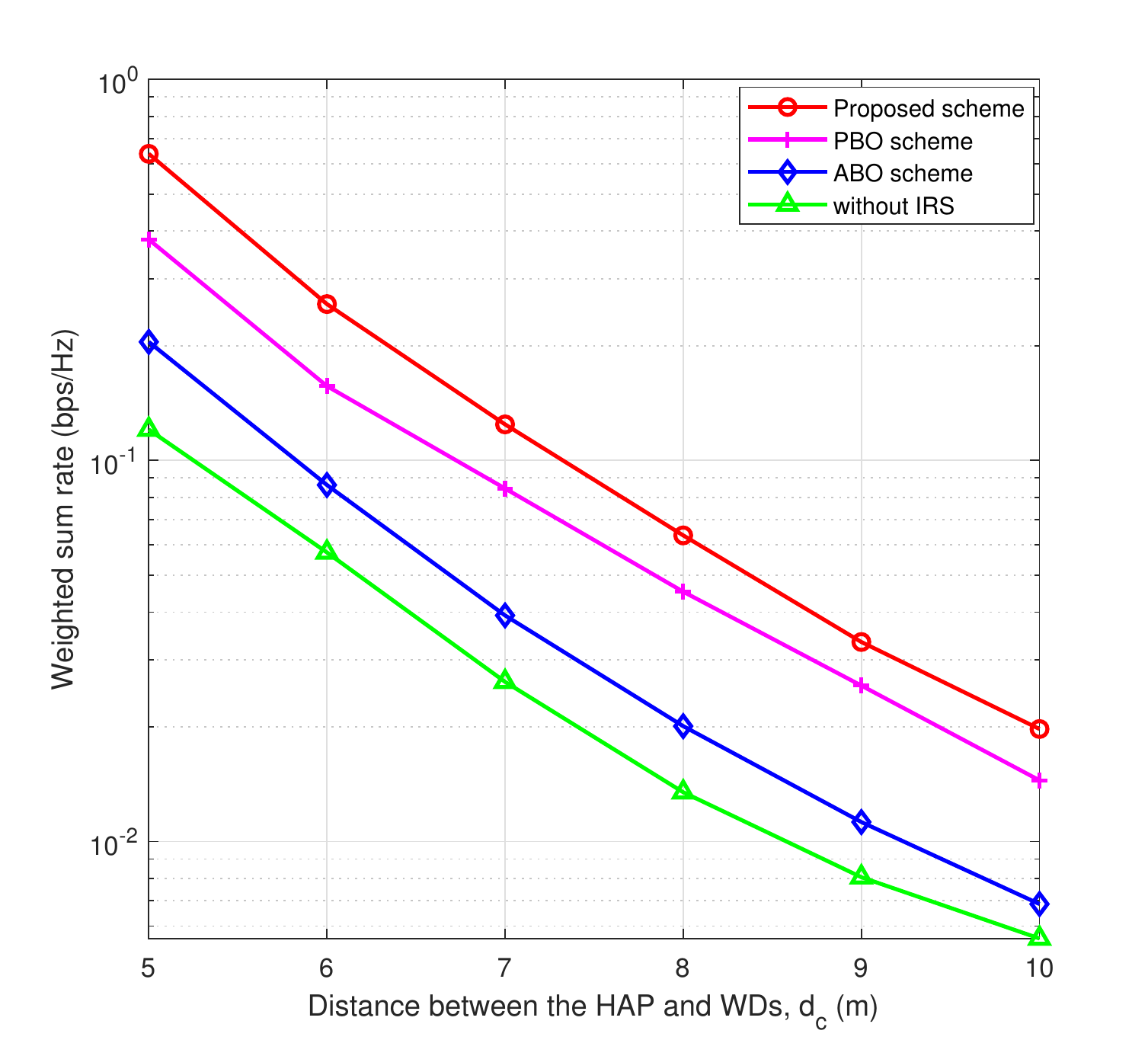}
   \end{center}\vspace{-.2in}
  \caption{The WSR performance versus the distance between the HAP and WDs.}
  \label{Fig.6}
\end{figure}
\begin{figure}
  \centering
   \begin{center}
      \includegraphics[width=0.48\textwidth]{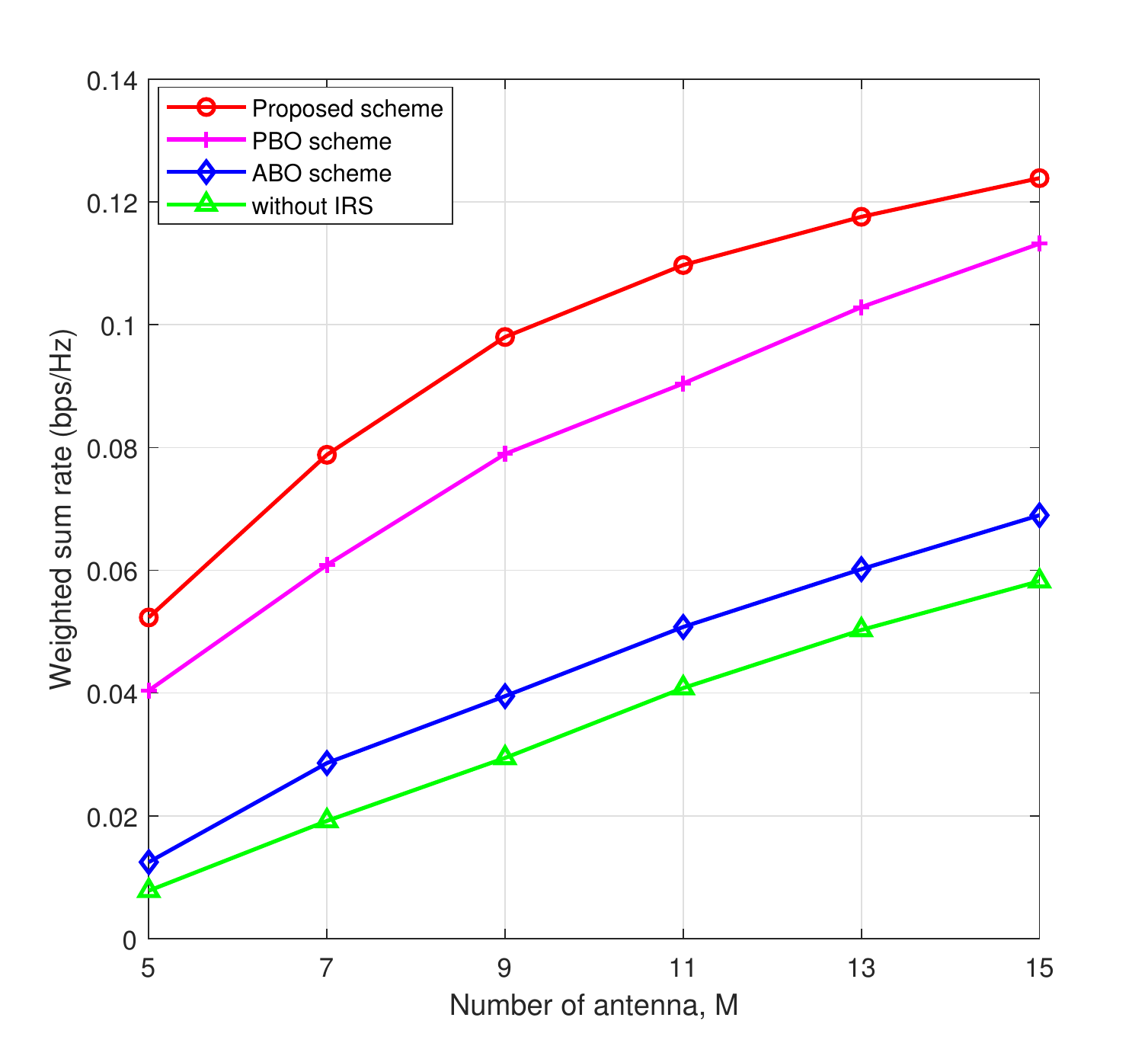}
   \end{center}\vspace{-.2in}
  \caption{The WSR performance versus the number of HAP antennas.}
  \label{Fig.7}
\end{figure}

Fig.~\ref{Fig.6} investigates the impact of the WDs deployment location to the WSR performance by varying $d_c$. We also see that the proposed scheme achieves evident performance advantages over the three benchmark methods. As expected, the performance gain of all the methods decreases as $d_c$ increases, because as the WDs move further away from both the HAP and IRS, and suffering from more severe signal attenuation in both energy harvesting and information transmission. The performance gain is especially evident when $d_c$ is large, e.g. $d_c> 9$ m, where the without-IRS scheme achieves very low rate (less than $10^{-2}$) while the proposed scheme still maintains relatively high rate (around ten times larger). This is because the WDs are unable to efficiently harvest sufficient energy for information transmission without the assistance of the IRS.

Then, we compare in Fig.~\ref{Fig.7} the WSR performance of all the schemes when the number of the HAP antennas (i.e., $M$) changes. It is observed that the WSR performance of all the methods increases with $M$ because of the higher spatial diversity gain. We also notice that our proposed scheme and the PBO scheme produce much better performance than the other two schemes due to the higher beamforming gain. Meanwhile, the performance of ABO scheme even performs close to the without-IRS scheme, which implies the importance of optimizing the passive beamforming to the throughput performance.

\begin{figure}
  \centering
   \begin{center}
      \includegraphics[width=0.48\textwidth]{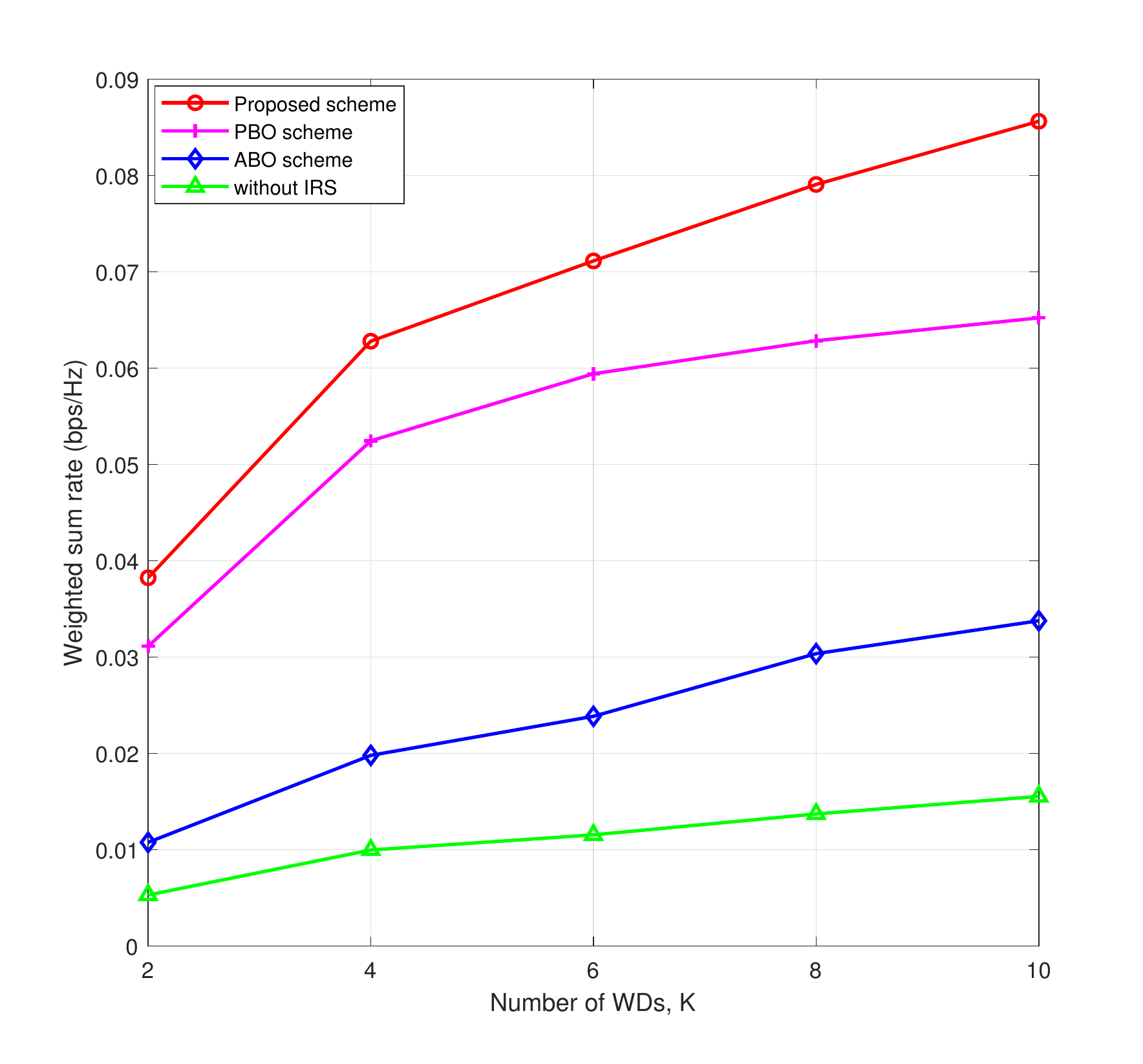}
   \end{center}\vspace{-.2in}
  \caption{The WSR performance versus the number of WDs.}
  \label{Fig.8}
\end{figure}

In Fig.~\ref{Fig.8}, we evaluate the WSR performance versus the number of WDs (i.e., $K$) for all the methods. Here, we vary $K$ from 2 to 10. It can be observed that the WSR performance increases with the number of WDs for all methods due to the benefit of multiuser diversity.  Meanwhile, the performance gap between the three IRS-assisted methods and the without-IRS scheme gradually increases with $K$.
This is because the uplink information transmission becomes interference-limited when the number of WDs is large. As a result, the optimal solution will allocate more time for transmitting information, which in consequence decreases the WET phase duration. In this case, the IRS becomes a critical factor that effectively increases the harvested energy of the WDs within the limited energy transfer time.

\begin{figure}
  \centering
   \begin{center}
      \includegraphics[width=0.48\textwidth]{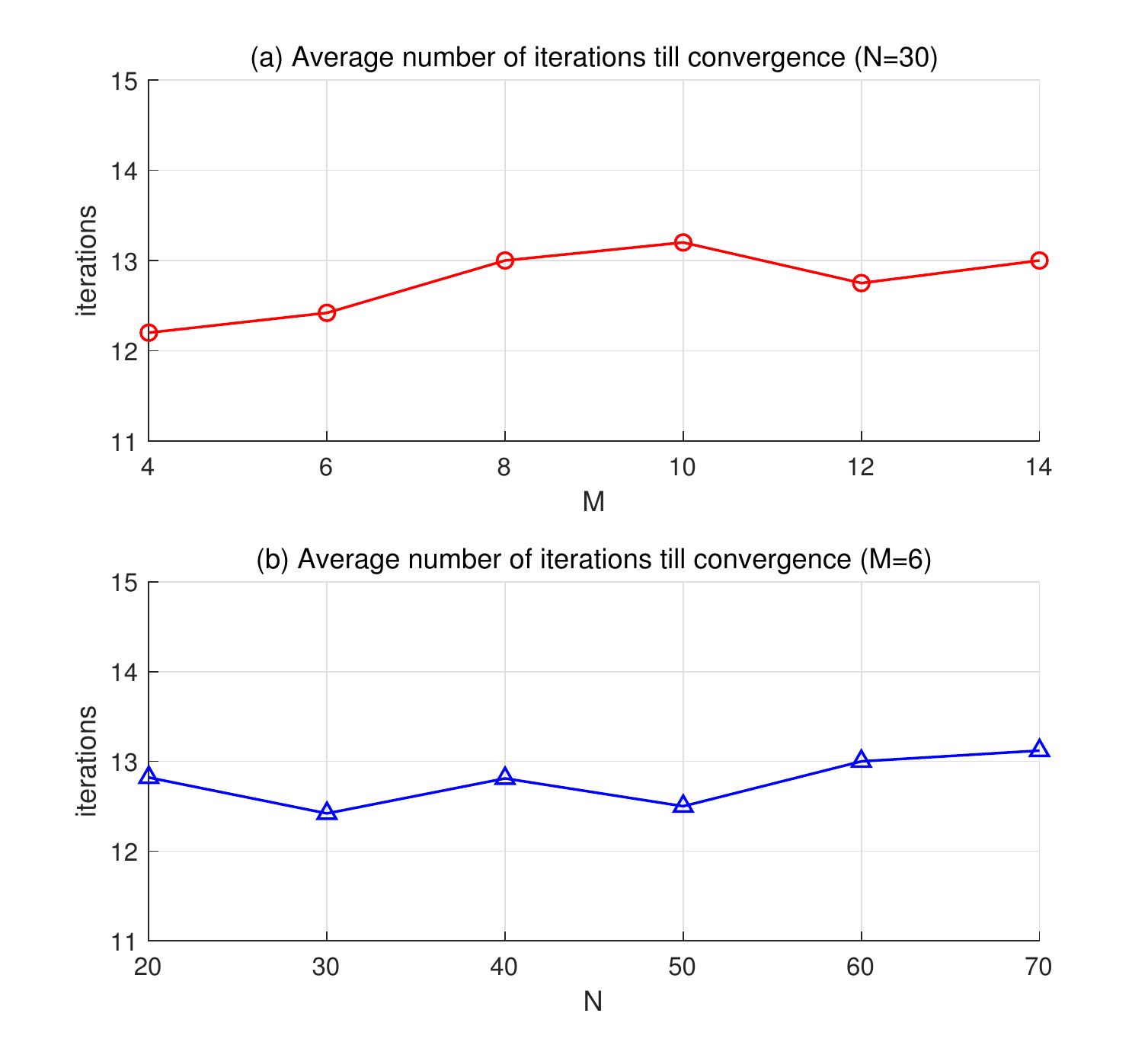}
   \end{center}\vspace{-.2in}
  \caption{The average iteration number of convergence of Algorithm 2, (a) as a function of $M$ under fixed $N = 30$; and (b) as a function of $N$ under fixed $M = 6$.}
  \label{Fig.9}
\end{figure}

We then show in Fig. 9 the convergence rate of Algorithm 2, for which the convergence is proved in Section IV.C. In particular, we plot the average number
of iterations required until the
algorithm converges under 100 independent simulations. Here, we investigate the convergence rate
when either the number of HAP antennas (i.e., $M$) or IRS reflecting elements (i.e., $N$) varies. With fixed $N=30$ in Fig. 9(a), we see that the number of iterations used till convergence does not vary significantly as $M$ increases. Similarly in Fig. 9(b), with a fixed $M=6$, we do not observe
significant increase of iterations when $N$ increases from 20 to
70. Besides, all the simulations performed in Fig. 9 require at most
20 iterations to converge. Therefore, we can safely estimate that
the number of iterations used till convergence is of constant
order, i.e., $\mathcal O(1)$. This indicates that the proposed method enjoys fast convergence even in a network with a large number of active antennas at the HAP or passive reflecting elements at the IRS.

To sum up, our simulation results show that the proposed joint beamforming and power control optimization achieves superior throughput performance in MISO WPCNs under various setups.  Meanwhile, we observe that, between the two better performing benchmark method, the PBO method outperforms the ABO scheme in all simulations. This indicates the importance of a refined passive beamforming design to achieve the high beamforming gain provided by the massive reflecting elements. Nonetheless, the significant performance gap between the PBO scheme and our proposed scheme confirms the benefit of joint active and passive beamforming optimization in enhancing the throughput performance of IRS-assisted WPCNs.

\section{Conclusions and Future Work}
In this paper, we have studied an IRS-assisted multiuser MISO WPCN. Specifically, the WSR optimization problem was formulated to jointly optimize the energy transmission time, the user transmit power, the active beamforming of the HAP and passive beamforming of IRS in both the UL and DL transmissions. To tackle this non-convex problem, we fixed the passive beamforming of the IRS and converted the original problem to an equivalent WMMSE problem, which was efficiently solved by a BCD method. Likewise, given user transmit power and active beamforming of the HAP, we optimized the passive beamforming of the IRS by the SDR technique.
This leads to a BSO-based iterative algorithm to update the two sets of variables alternately.
At last, we applied an one-dimensional search method to obtain the optimal WET time. By comparing with representative benchmark methods, we showed that the proposed joint optimization achieves significant
performance advantage and effectively enhances the throughput performance in multi-user MISO WPCNs under different practical network setups.

Finally, we conclude the paper with some interesting future working directions. First, it is interesting to consider a practical non-linear energy harvesting model, such that the active and passive beamforming design in the DL energy transfer must be adapted to improve the energy harvesting efficiency of all users. In addition, it is also promising to consider more realistic imperfect CSI case, where the knowledge of the cascaded HAP-IRS-WD channels are under uncertainty due to channel estimation error. To tackle the problem, we may investigate the robust transmission design for IRS-assisted MISO communication systems under a stochastic CSI error model. Moreover, although some recent works have considered IRS with only a finite number of phase shifts at each element. In this case, the beamforming problem becomes very challenging due to the combinatorial phase shift variables and the strong coupling
with the other system design parameters. One possible way is to introduce a learning-based discrete beamforming method for reducing the computational complexity. At last, it is also challenging to extend the considered network model to other practical setups, such as full-duplex transmission, cluster-based cooperation, hardware-constrained reflection at the IRS, and non-interference scenario, etc.

\section*{Appendix 1\\Proof of Proposition 1}
\emph{Proof:} Firstly, by employing the well-known rate-MMSE equivalence established in\cite{2011:Shi}, we have
\begin{equation}
\label{rmm}
\log(1+\gamma_i)=\log([e_i^{\text{MMSE}}]^{-1}),
\end{equation}
where $e_i^{\text{MMSE}}$ denotes the MMSE of received signal $s_i$ from WD$_i$, and it is expressed as
\begin{equation}
e_i^{\text{MMSE}}=\min\ e_i,
\end{equation}
where $e_i$ is defined as (\ref{ei1}). Then, by substituting it into (\ref{rmm}), we have
\begin{equation}
\label{gam}
\log(1+\gamma_i)=\log([\min\ e_i]^{-1})=\max\ \log({e_i}^{-1}).
\end{equation}

Consider the following equality
\begin{equation}
\log(x^{-1})=\max\limits_{y\ge0} \ (\log (y)-(xy)+1),
\end{equation}
where the optimal solution is achieved at $y^*=x^{-1}$. Thus, we rewrite (\ref{gam}) as
\begin{equation}
\begin{aligned}
\log(1+\gamma_i)&=\max\ \log({e_i}^{-1})\\
&=\max\limits_{q_i\ge0}\ \left(\log (q_i)-(q_ie_i)+1\right)\\
&=\min\limits_{q_i\ge0}\ \big((q_ie_i)-\log (q_i)-1\big).
\end{aligned}
\end{equation}
Accordingly, given $t=\bar t$, problem (P1) can be transformed into the equivalent problem as (P2).
\section*{Appendix 2\\Proof of Proposition 2}
\emph{Proof:} 
%
The Karush-Kuhn-Tucker (KKT) conditions of (P3) with respect to $\mathbf W^*$ are
\begin{equation}
\label{bara}
\bar{\mathbf A} \mathbf W^*=\mathbf 0,
\end{equation}
\begin{equation}
\label{suc}
\mu_i^*\ge0,\mu_0^*\ge0,\mathbf W^*\succeq 0,\forall i\in\mathcal K,
\end{equation}
\begin{equation}
\label{mu0}
\mu_0^*\Big(\text{tr}(\mathbf W^*)-P_0\Big)=0,
\end{equation}
\begin{equation}
\label{mui}
\mu_i^*\Big((1-\bar t)P_i+E_i^{(2)}-\eta\bar t\text{tr}(\boldsymbol b_i\boldsymbol b_i^H\mathbf W^*)\Big)=0,\forall i\in\mathcal K,
\end{equation}
where $\bar{\mathbf A}=\sum_{i\in\mathcal K}\mu_i^*\eta\bar t\boldsymbol b_i\boldsymbol b_i^H-\mu_0^* \mathbf I$.

In practice, we always find a rank-one energy beamforming matrix by using the derived optimal conditions in (\ref{bara})-(\ref{mui}). We first consider the case of $\mu_i^*= 0$ and $\mu_0^*>0$. In this case, we have $\mathbf W^*=\mathbf 0$ from (\ref{bara}) since $\bar{\mathbf A}=-\mu_0^* \mathbf I$, which contradicts the complementary slackness condition (\ref{mu0}). Also, for the case where $\mu_i^*>0$ and $\mu_0^*=0$, $\mathbf W^*=\mathbf 0$ from (\ref{bara}) since $\bar{\mathbf A}=\sum_{i\in\mathcal K}\mu_i^*\eta\bar t\boldsymbol b_i\boldsymbol b_i^H$, which contradicts the complementary slackness condition (\ref{mui}).
Hence, both $\mu_0^*$ and $\mu_i^*$ are greater than zero, i.e., $\mu_0^*>0$ and $\mu_i^*>0$,  $\forall i\in \mathcal K$.

Next, we denote $\mathbf B=\sum_{i\in\mathcal K}\mu_i^*\eta\bar t\boldsymbol b_i\boldsymbol b_i^H$. Let the eigenvalue decomposition of matrix $\bar {\mathbf A}$ be $\bar {\mathbf A}=\mathbf U(\boldsymbol\Lambda-\mu_0^* \mathbf I)\mathbf U^H$, where $\mathbf U\in\mathbb C^{M\times M}$ and $\boldsymbol\Lambda=\text{diag}(\lambda_1,\cdots,\lambda_M)\in\mathbb C^{M\times M}$ with $\lambda_1\ge\cdots\ge\lambda_M$ are the eigenvector matrix and eigenvalue matrix of $\mathbf B$, respectively. Since $\mu_i^*>0$, $\forall i\in \mathcal K$, $\mathbf B$ is always a positive semidefinite, and resulting in the non-negative eigenvalues $\lambda_j$, for $j=1,\cdots,M$. Thus, for $\bar {\mathbf A}$ to have non-positive eigenvalues, i.e.,  $\lambda_j-\mu_0^*\le 0$, we obtain $0\le\lambda_j\le\mu_0^*$. When $\mu_i^*>0, \forall i\in \mathcal K$, we have $\text{rank}(\mathbf B)>1$, the maximum eigenvalue $\lambda_1>0$.

 Note that if $\mu_0^*>\lambda_1$, $\bar{\mathbf A}$ becomes a full-rank and negative-definite matrix. Thus, we obtain $\mathbf W=\mathbf 0$ from (\ref{bara}), which contradicts the complementary slackness condition (\ref{mu0}) since $\mu_0^*>0$. Therefore, we obtain the optimal dual variable $\mu_0^*$ as $\mu_0^*=\lambda_1$. We define $\bar{\mathbf A}\mathbf u_1=\mathbf 0$, where $\mathbf u_1$ is the unit-norm eigenvector of ${\mathbf B}$ corresponding to the maximum eigenvalue $\lambda_1$. From (\ref{bara}) and (\ref{suc}), we obtain the optimal $\mathbf W^*=\epsilon \mathbf u_1\mathbf u_1^H$ for any $\epsilon\ge0$. Next, we find $\epsilon$ from (\ref{mu0}), i,e., $P_0-\text{tr}(\mathbf W^*)=0$ due to $\mu_0^*>0$, which leads to $\text{tr}(\mathbf W^*)=\epsilon=P_0$. $\hfill\blacksquare$

\end{document}